\documentclass[aps, twocolumn]{revtex4}

\usepackage{float}
\usepackage{dcolumn}
\usepackage{amsmath}
\input{epsf}
\input{epsfx}

\ifx\pdftexversion\undefined
  \usepackage{graphicx}
\else
  \usepackage[pdftex]{graphicx}
\fi
\usepackage{epsfig}
\usepackage{ulem}

\begin{document}
\title{Local tuning of photonic crystal cavities using chalcogenide glasses}

\author{Andrei Faraon \footnote[1] {Electronic address: faraon@stanford.edu }, Dirk Englund}
\affiliation{Department of Applied Physics, Stanford University, Stanford, CA, 94305, USA}

\author{Douglas Bulla, Barry Luther-Davies}
\affiliation{Centre for Ultrahigh-bandwidth Devices for Optical Systems (CUDOS), Laser Physics Centre, Australian National University, ACT, 0200, Australia}

\author{Benjamin J. Eggleton}
\affiliation{Centre for Ultrahigh-bandwidth Devices for Optical Systems (CUDOS), School of Physics, University of Sydney, NSW, 2006, Australia}

\author{Nick Stoltz, Pierre Petroff}
\affiliation{Department of Electrical and Computer Engineering, University of California, Santa Barbara, CA 93106}

\author{Jelena Vu\v{c}kovi\'{c}}
\affiliation{E. L. Ginzton Laboratory, Stanford University, Stanford, CA, 94305, USA}


\begin{abstract}

We demonstrate a method to locally change the refractive index in planar optical devices by photodarkening of a thin chalcogenide glass layer deposited on top of the device. The method is used to tune the resonance of GaAs-based photonic crystal cavities by up to 3 nm at 940 nm, with only 5\% deterioration in cavity quality factor. The method has broad applications for post-production tuning of photonic devices.
\end{abstract}

\maketitle

On-chip integration of optical components promises to greatly enhance speed and reduce costs in optical communications applications, such as interconnects and optical logic. Photonic crystal (PCs) devices are one of the most promising platforms for on-chip integration, as they can combine optical waveguides, resonators, dispersive devices, lasers or modulators \cite{Notomi_PC_devices_review,Fatih_stopped_light,Notomi_Silicon_switch}. Such devices can be patterned with existing semiconductor lithographic techniques. However, they are highly sensitive to fabrication imperfections \cite{Dirk_direct_PC_analysis} and a practical method to locally tune their optical properties is needed. In this paper we present a method for tuning GaAs PC devices, based on chalcogenide glasses. Chalcogenide glasses quasi-permanently change their optical properties when illuminated with light above their band gap, and have been used to tune optical devices as quantum cascade lasers \cite{QuantCascLaserTuning} The tuning of PCs devices directly fabricated in chalcogenide glasses has already been shown in Ref.\cite{Egg_wg_posttune}, but many other applications rely on PC fabricated in other materials such as group IV and III-V semiconductors. One such application is quantum information with InAs quantum dots (QDs) embedded in GaAs photonic crystal structures \cite{AndreiWgCoupling,AndreiTtune, DirkAndreiSPhtransfer}. The performance of these devices relies on precise wavelength-matching of the cavity to embedded QDs. One way to achieve spectral matching is by temperature tuning, which shifts the cavity and QD at different rates so that they can be made to intersect if the original mismatch is small\cite{AndreiTtune}.  However, it is often necessary to shift the cavity independently as well, especially in photonic networks\cite{DirkAndreiSPhtransfer}. To date, several non-local techniques have been developed to control cavity wavelength (see Ref.\cite{HennessyDigitalEtch, StraufMonolayerCavityTuning}), and a local tuning method based on tuning the resonance by bringing a nanowire in the proximity of the photonic crystal cavity has been reported by Grillet et. al.  in Ref.\cite{Grillet_taper_tune}. However, none of these techniques is reliable for independent on-chip tuning of large ensembles of photonic crystal cavities.

\begin{figure}[htbp]
    \includegraphics[width=3.5in]{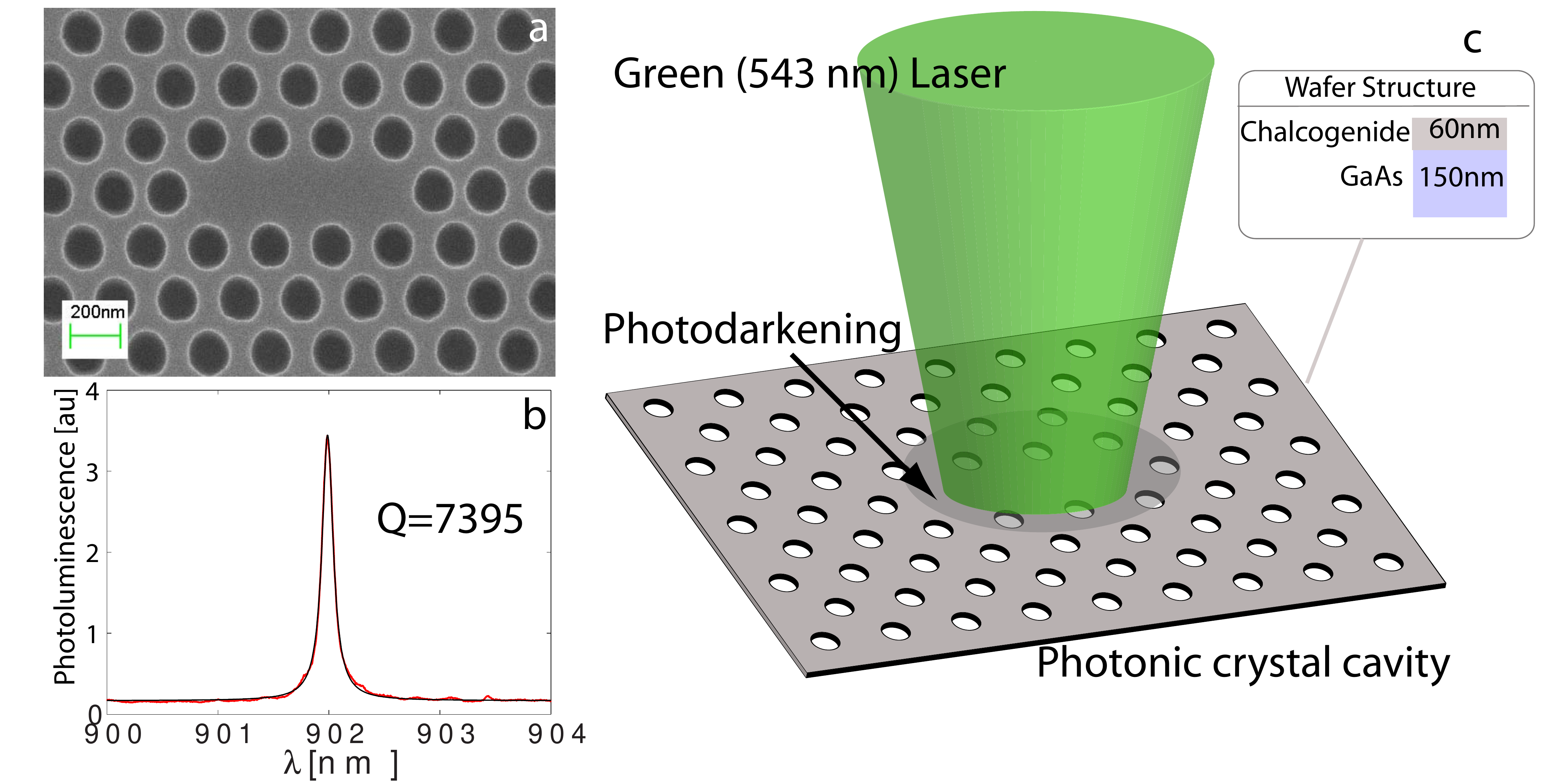}
\caption{(a) Scanning electron microscope image of the photonic crystal cavity fabricated in GaAs before the deposition of As$_{2}$S$_{3}$. (b) Cavity spectrum before the chalcogenide deposition indicating a quality factor Q=7395. (c)Schematic of the method for local cavity tuning. A layer of As$_{2}$S$_{3}$ is deposited on top of the photonic crystal cavity. Then a laser tuned close to the As$_{2}$S$_{3}$ band gap is focused on the cavity, increasing the effective refractive index and causing a resonance red-shift.}
\label{fig:cavity}
\end{figure}

In our approach, a photosensitive chalcogenide glass layer is deposited on prefabricated GaAs/InAs devices. Linear three-hole defect \cite{NodaL3} PC cavities were first fabricated in a 150 nm thick GaAs membrane containing a central layer of InAs quantum dots (QDs) as described in  Ref.\cite{05PRLEnglund}. Arsenic trisulphide films with thickness between 30 nm and 100 nm were deposited onto the photonic crystals using thermal evaporation from a temperature controlled baffled boat in a chamber pumped to a base pressure of $3 \times 10^{-7}$ torr. The deposition geometry was chosen to ensure that the flux of material struck the sample close to normal incidence to prevent coating the inside of the holes. To improve film adhesion the sample surfaces were cleaned using 50 eV Ar ions prior to deposition.

Thermal evaporation results in films with substantially different bond structure from the bulk glass. The films have been found to contain disconnected molecular cage-like structures \cite{Chalc_schulte} that form in the vapour phase and are then frozen into the deposited film. These cage-like structures are, however, meta-stable and can``open" with optical excitation near the band edge or by heating that allows re-bonding to occur which results in polymerization of a more extended glass network. This re-bonding process is accompanied by an increase in the refractive index and a small decrease of the material volume \cite{KuglerChalcModel}. The thermally evaporated films had an index (at 1550 nm) of $ \sim 2.31$ compared with the bulk glass whose index is ‰2.43. After deposition the films were partially polymerized  by annealing at 130$^{\circ}$C for 24 hours prior to use which increased the refractive index to $\sim 2.38$. Subsequent irradiation of the films with actinic light at high fluence can increase the film index to the bulk value. 


The experiment was performed at cryogenic temperature (less than $\sim 60K$) to obtain luminescence from the embedded InAs quantum dots, as needed for quantum information processing applications.. This illustrates that the method works at low temperatures, though we stress that it is applicable to room temperature nanophotonic circuits. The sample was placed inside a continuous-flow liquid helium cryostat at 10K and the QD photoluminescence was used to measure the cavity resonance. A confocal microscope setup and a laser tuned at 780 nm excited quantum dot luminescence while a spectrometer monitored the signal. A 543 nm HeNe laser ($1 \mu W$) focused to $\sim 1\mu m ^{2}$ through the same confocal setup was used for photodarkening of the As$_{2}$S$_{3}$ layer (Fig.\ref{fig:cavity}). This wavelength was chosen because it is close to the 527 nm bandgap of As$_{2}$S$_{3}$.


The thickness of the As$_{2}$S$_{3}$ influences both the quality factor of the cavity and the maximum tuning range. For this reason we experimented with three different thicknesses: 30, 60 and 100 nm (samples S30, S60 and S100). For each sample, the spectrum of the cavities was recorded before and after the deposition of the chalcogenide layer. For samples S60(S30), the deposition caused the quality factor to degrade by $\sim 5\%$($30\%$) from an average value of $\sim 8500$($10000$)  while the resonant wavelength shifted by $\sim \text{40 nm}(\text{28 nm}$). For sample S100 the Q degradation was more severe, from $\sim 6500$ to $\sim 1000$ and for this reason we mainly concentrate on samples S30 and S60.  

\begin{figure}[htbp]
    \includegraphics[width=3.5in]{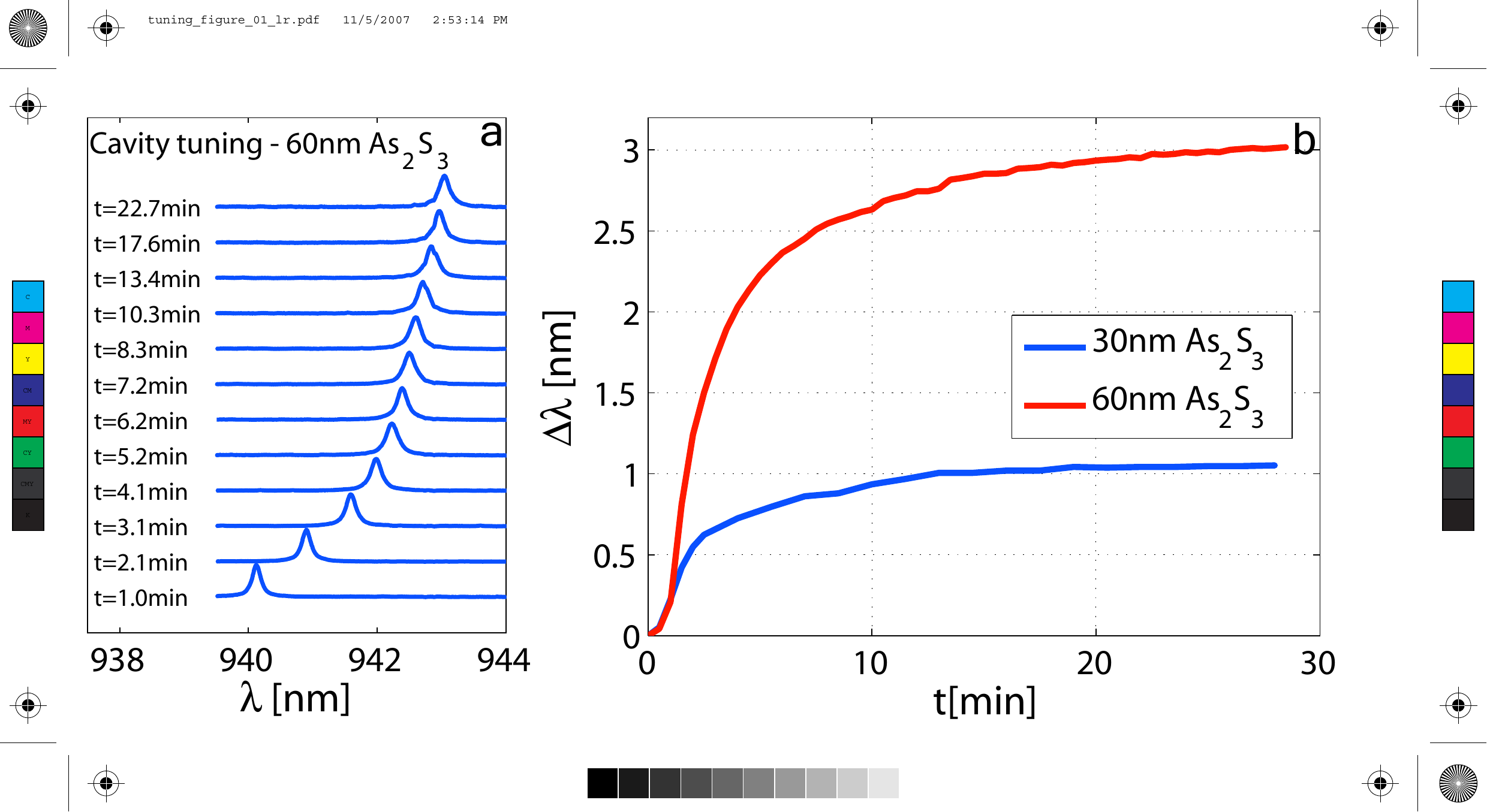}
\caption{(a)Spectra showing the shift of the cavity resonance because of the photodarkening of the 60 nm thick chalcogenide layer. (b) Time dependence of the cavity resonance for 60 nm and 30 nm As$_{2}$S$_{3}$ during the tuning process.}
\label{fig:spectra-time}
\end{figure}

With the chips mounted in the cryostat, we focused the 543 nm laser on the PC cavities for a fixed time and recorded the cavity spectrum. For sample S60, the cavity resonance shifted by up to 3 nm as shown in Fig.\ref{fig:spectra-time}(a). For $1 \mu W$ of green laser power focused on a spot size of $\sim 1 \mu m ^{2}$, the cavity tuning rate levels off after about 20 minutes, as shown in Fig.\ref{fig:spectra-time}(b). This saturation time is inversely proportional to the energy flux incident on the sample surface. During the tuning process the quality factor degraded by 20\%. The maximum tuning range dependends on the thickness of the chalcogenide layer. For 30 nm and 100 nm, a tuning range of 1 nm (Fig.\ref{fig:spectra-time}(b)) and 4 nm was observed, respectively. The change of the cavity resonance was stable after the green laser was turned off.

Illumination of As$_{2}$S$_{3}$ with light at 543 nm causes changes both in the refractive index and in the density of the material. Experiments at room temperature with films of As$_{2}$S$_{3}$ show that the increase in the refractive index is accompanied by a $\sim 1.5\%$ decrease in the film thickness. The decrease in thickness should result in a blue-shift of the cavity resonance. The red-shift observed experimentally implies that the dominant effect responsible for the shift of the cavity resonance is the change in the refractive index. We note that during the cool-down and exposure steps, strain can build up between the glass and substrate that slightly shifts QD emission lines.  After an initial strain-induced shifting early in the exposure, we find that QDs are nearly stationary for the majority of the tuning range, presumably because of nanofractures that relieve the strain.  We are currently investigating improved methods for resetting strain.

For our experiment, the smallest area that can be locally tuned is limited by the focus size of the laser beam ($\sim 1\mu m ^{2}$). The locality of the technique allows for independent tuning of interconnected optical components on photonic crystal chips. The method is not only suitable for GaAs devices, but can possibly be implemented with any other materials, including silicon nanophotonic circuits. Also, the As$_{2}$S$_{3}$ can easily be replaced by other types of chalcogenide glasses or other photosensitive materials depending on the specific application.


In conclusion, we have shown that As$_{2}$S$_{3}$ can be combined with semiconductor photonic crystals to create nanophotonic devices whose optical properties can be independently fine-tuned on the same chip. This technique is relevant for fabrication of integrated nanophotonic circuits for classical and quantum information processing, including applications such as filtering, multiplexing, optical storage, fine-tuning of modulators and lasers, and local tuning of distinct PC cavities on GaAs/InAs chips for quantum optics.


	Financial support was provided by ONR Young Investigator Award, the MURI Center for photonic quantum information systems (ARO/DTO program No. DAAD19-03-1-0199)  and NSF Grant No. CCF-0507295. Work was performed in part at the Stanford Nanofabrication Facility of NNIN supported by the National Science Foundation under Grant ECS-9731293. CUDOS is an Australian Research Council Centre of Excellence.


\bibliographystyle{unsrt}

\end{document}